\documentclass[dvips]{article}

\usepackage{icrc2011}

\title{Observations of the Crab pulsar with the MAGIC telescopes}

\newcommand{\etal}{\MakeLowercase{\textit{et al. }}} 
\shorttitle{T.Y. Saito \etal Observations of the Crab Pulsar with MAGIC}

\authors{T.Y. Saito$^{1}$, M. L\'opez$^{2}$, G. Giavitto$^{3}$, S. Klepser$^{3}$, T. Schweizer$^{1}$,R. Zanin$^{3}$\\ on behalf of the MAGIC collaboration}
\afiliations{
$^1$Max-Planck-Institut f\"ur Physik, M\"unchen, Germany\\ 
$^2$Universidad Complutense, Madrid, Spain\\
$^3$Institut de Física d'Altes Energies, Barcelona, Spain 
 }
\email{tysaito@mppmu.mpg.de}

\abstract{
We report on the observations of the Crab pulsar with the MAGIC telesopes.
Data were taken both in the mono-mode ($>25$ GeV) and in the stereo-mode ($>50$ GeV). Clear signals from the two peaks were detected with both modes and
the phase resolved energy spectra were calculated. By comparing 
with the measurements done by Fermi-LAT, we found that the energy spectra
of the Crab pulsar does not follow a power law with an exponential cutoff,
but that it extends as a power law after the break at around 5 GeV.
This suggests that the emission above 25 GeV is not dominated by the
curvatura radiation, which is inconsistent with the standard prediction
of the OG and SG models.

}
\keywords{pulsar, the Crab pulsar, MAGIC}

\begin{document}
\maketitle
\section{Introduction}
The Crab nebula is the compact object left over after a historic supernova explosion that occurred 
in the year 1054 A.D.. 
The pulsar B0531+21  (also commonly named Crab pulsar) is located at its center, and emits
 strong pulsed radiation in a wide energy range from radio to high energy gamma-rays.
\\
The Crab pulsar and a few other pulsars are among the brightest known sources at 1 GeV.
However, a spectral steepening made
their detection above 10 GeV elusive despite numerous efforts. 
The energy thresholds of imaging atmospheric Cherenkov telescopes (IACTs) were, in general,
too high, while the gamma-ray collection area of satellite-borne detectors were
too small to detect pulsars above 10 GeV. On the other hand, a precise measurement of the energy spectrum at and above the steepening 
leads to an important verification for the standard pulsar models. 
In the case of the Polar Cap (PC) 
model\cite{PCmodel} ,
so-called super-exponential cutoff is expected, while the Outer Gap (OG) model \cite{OGmodel} and Slot Gap (SG) model \cite{SGmodel}, predict
a clear exponential cutoff. The highest energy of the detected photons can be directly 
converted to the lower limit on the distance of the emission region from the stellar surface,
which should be a few times the stellar radius according to the PC model.

In 2008, the MAGIC telescope detected the Crab pulsar above 25 GeV \cite{CrabScience}
with the newly implemented
trigger system, the Sum trigger \cite{SumT}. This detection excluded the PC model.
In August 2008, the new satellite borne gamma-ray detector with 1 m$^2$ collection area, Fermi-LAT, 
became operational and it could measure the spectra of gamma-ray pulsars up to a few tens of GeV.
The spectra measured by Fermi-LAT could be described with a power law with an exponential cutoff,
which also rejected the polar cap model and supported the OG and the SG model. \\

However, the cutoff energy of the Crab pulsar spectrum determined by Fermi-LAT was $\sim 6$ GeV,
while MAGIC detected the signal above 25 GeV. In order to verify the exponential cutoff 
spectrum, i.e. OG and SG models, the precise comparison of the energy spectra measured by the two instruments
is needed. Here we present the spectral study of Crab pulsar, using the public Fermi-LAT
 data and four years of MAGIC data recorded by the single telescope and
 the stereoscopic system.

\section{MAGIC telescope}
The MAGIC telescope is a new generation IACT
 located on the Canary island of
 La Palma (27.8$^\circ$ N, 17.8$^\circ$ W, 2225 m asl). 
It consists of two telescopes with a reflector diameter of 17 m.
The first telescope was build in 2002-2003 and have been operational since 2004.
Thanks to the world largest reflector, the energy threshold of the first MAGIC telescope
with the standard trigger is 60 GeV, that is the lowest among IACTs.
In order to detect gamma-ray pulsars, the new trigger system called Sum trigger was 
developed and implemented in October 2007. It reduced the energy threshold further,
down to 25 GeV, which resulted in the detection of the Crab pulsar \cite{CrabScience}.
In 2009, the second MAGIC telescope was build $\sim 80$ m apart from the first telescope.
The second one is basically a clone of the first one, while the Sum trigger system
is not yet implemented to it. 

We observed the Crab pulsar with the stereoscopic mode from 2009.
The stereo trigger requires a coincidence of the triggers of both telescopes.
For a technical reason, the Sum trigger in the first telescope 
cannot participate in the stereo trigger,
i.e., stereoscopic observations were based on the standard trigger for both telescopes.
The energy threshold of the stereo mode is about 50 GeV.

\section{Mono-mode observations}

MAGIC observed the Crab pulsar with a single telescope with the Sum trigger
in winter 2007-2008 and winter 2008-2009.
After the careful data selection, total effective observation time was 25 hours and 34 hours
for the first and the second campaign, respectively. The energy threshold of
these observations are 25 GeV. 
Normally, IACT technique utilizes many image parameters to distinguish between 
hadron events and gamma-ray events. However, in the case of mono-mode observations
at the very low energy regime below 60 GeV,
the image parameter are almost powerless except for the Hillas parameter ALPHA. 
Therefore, the hadron background rejection was done only based on ALPHA.
 
\begin{figure}[h]
\centering
\includegraphics[width=3.in]{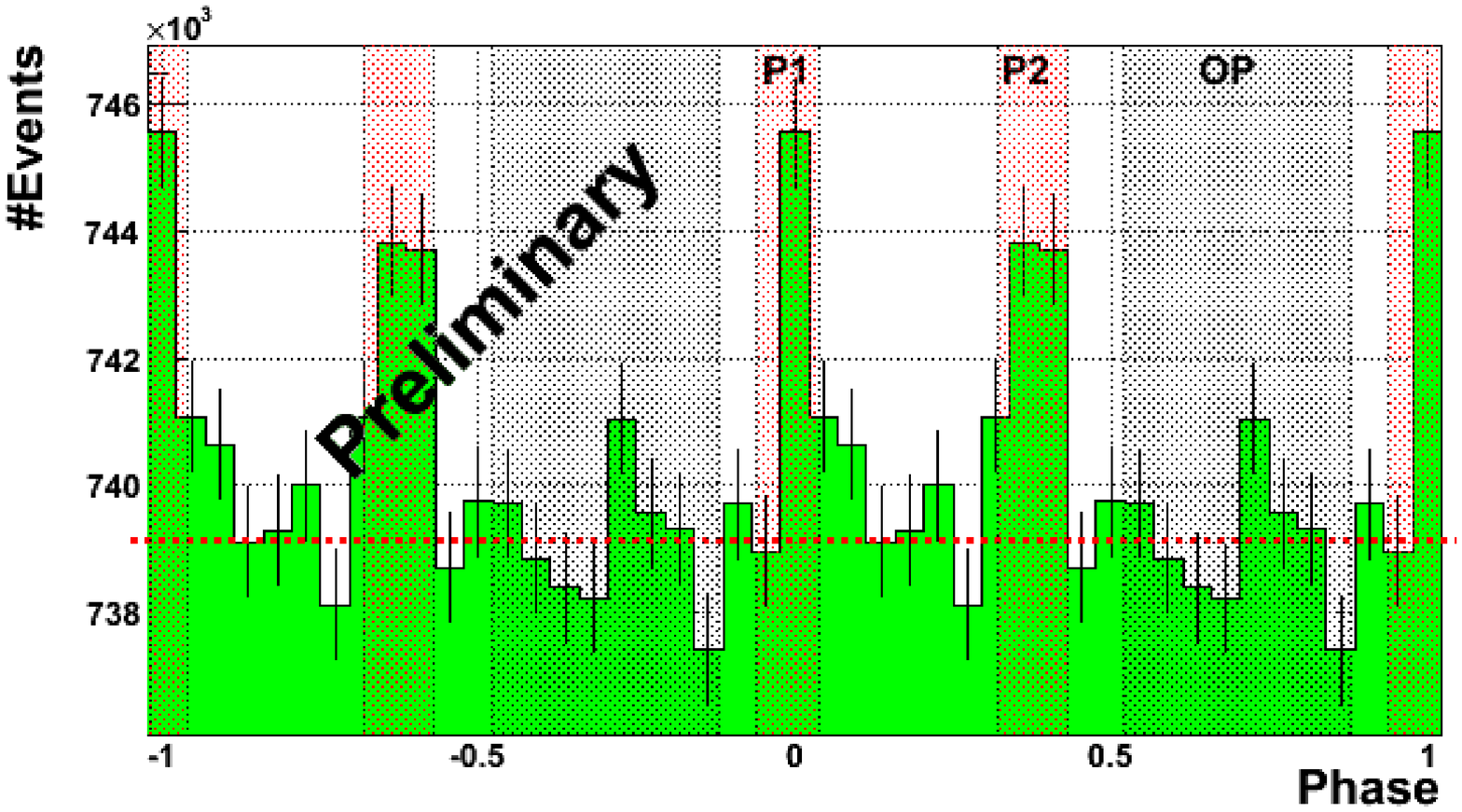}
\includegraphics[width=3.in]{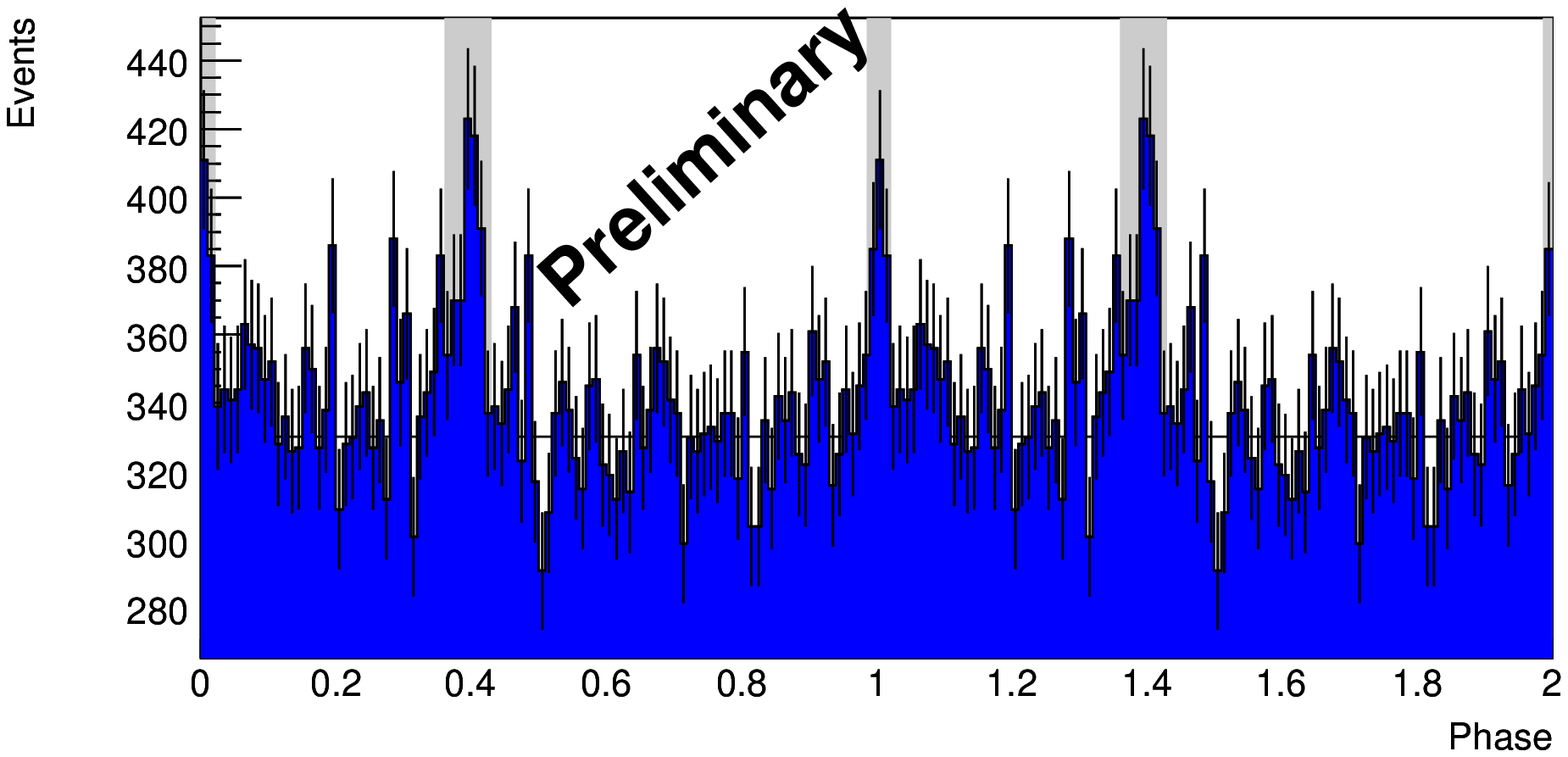}
\caption{The light curve of the Crab pulsar obtained by MAGIC
above 25 GeV  with the mono-mode (top) and above 50 GeV with the stereo mode (bottom). The effective observation time is  59 hours for mono
and 73 hours for stereo. 
} \label{FigMonoLC}
\end{figure}

 The light curve of the Crab pulsar obtained with the mono-mode observation is 
shown in the upper panel of Fig. \ref{FigMonoLC}. Following the usual convention
 \cite{Fierro1998} of 
P1 (phase interval -0.06 to 0.04) and P2 (0.32 to 0.43), the numbers of excess events in P1 and P2
are $6200 \pm 1400$ (4.3 $\sigma$) and $11300 \pm 1500$ ($7.4\sigma$).
By summing up P1 and P2, the excess corresponds to 7.5$\sigma$. 
The background level was estimated by using the so-called Off Pulse phase (0.52 - 0.88)
\cite{Fierro1998}.
 
\begin{figure*}[h]
\centering
\includegraphics[width=2.in]{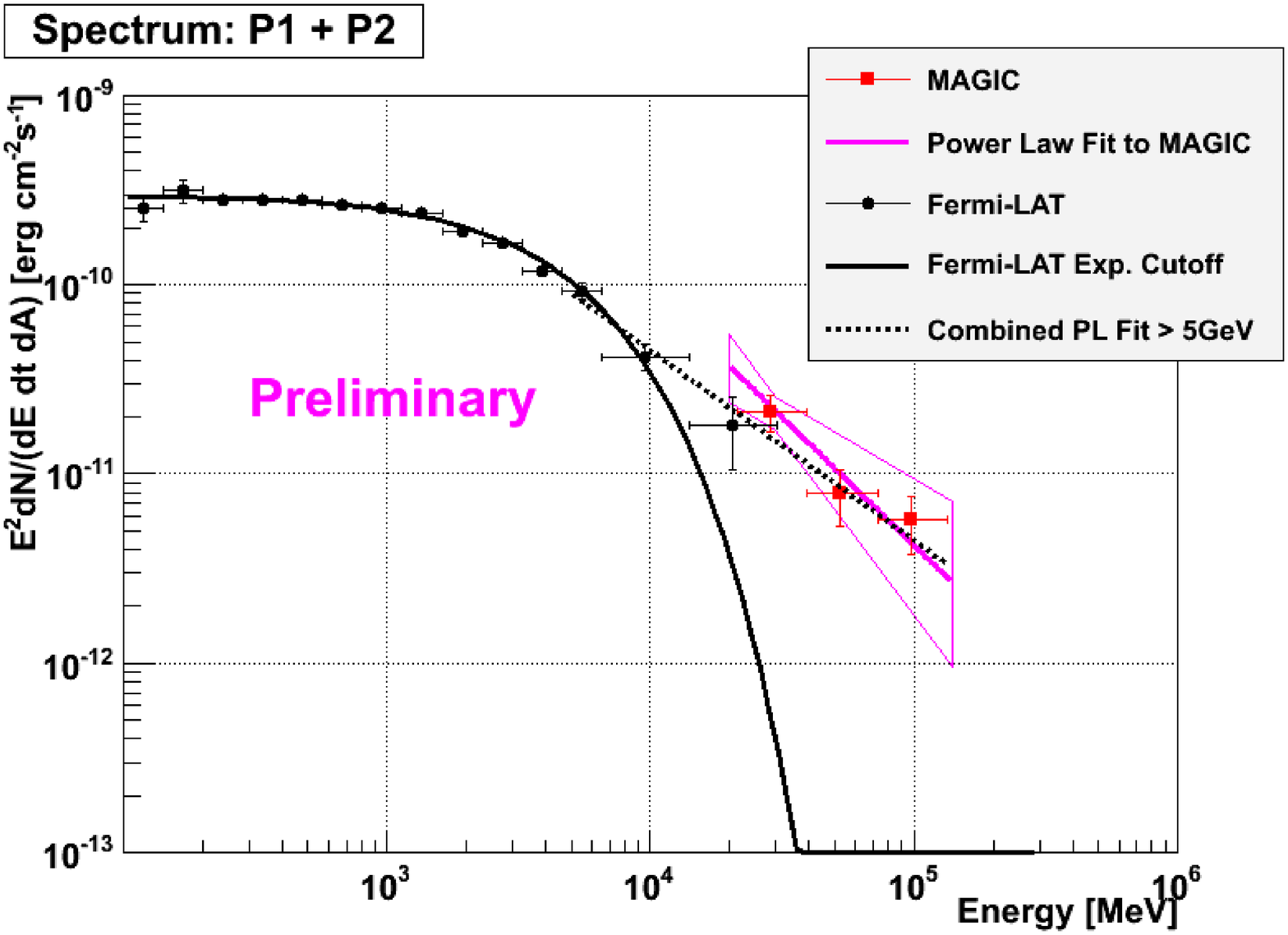}
\includegraphics[width=2.in]{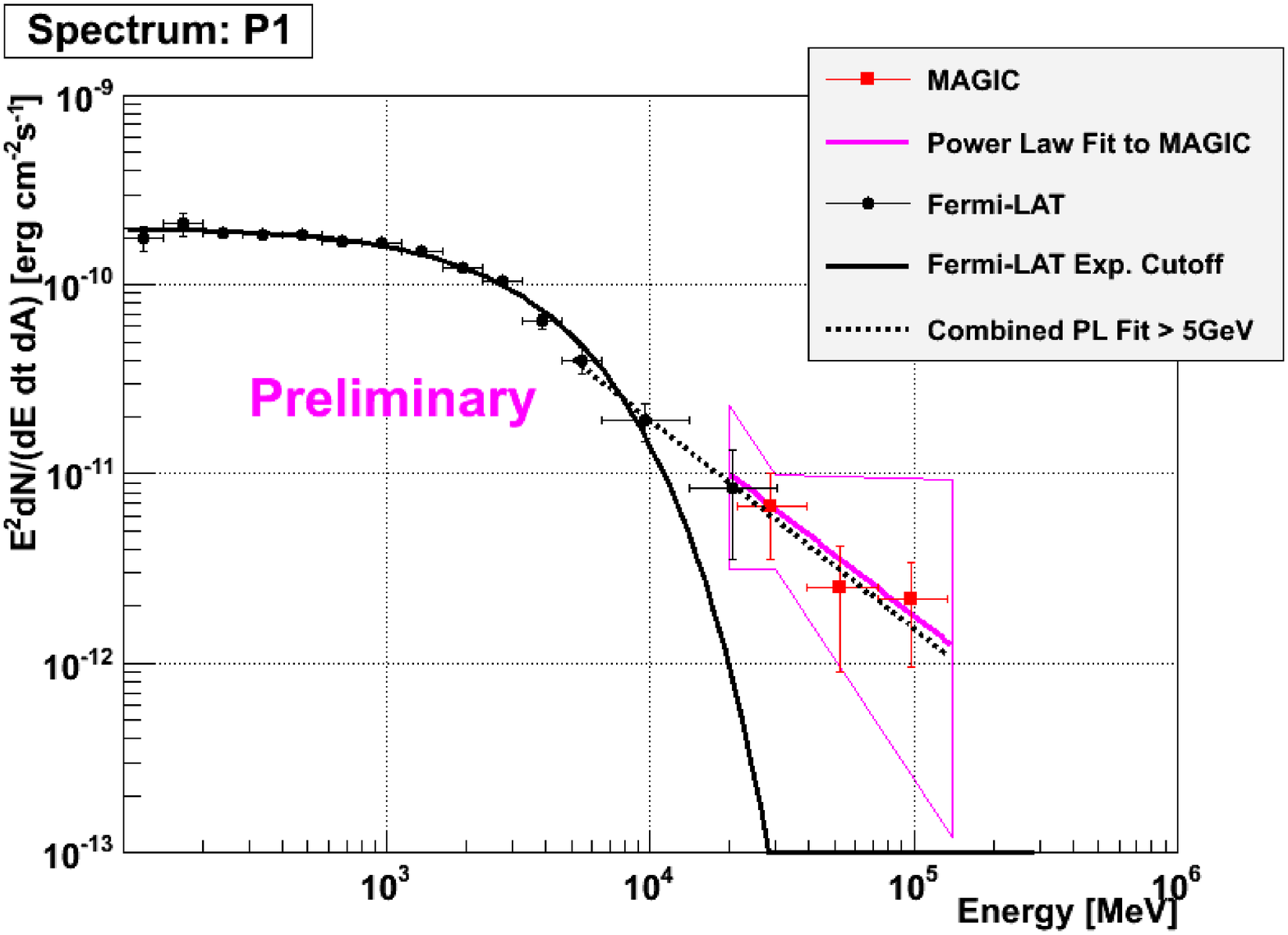}
\includegraphics[width=2.in]{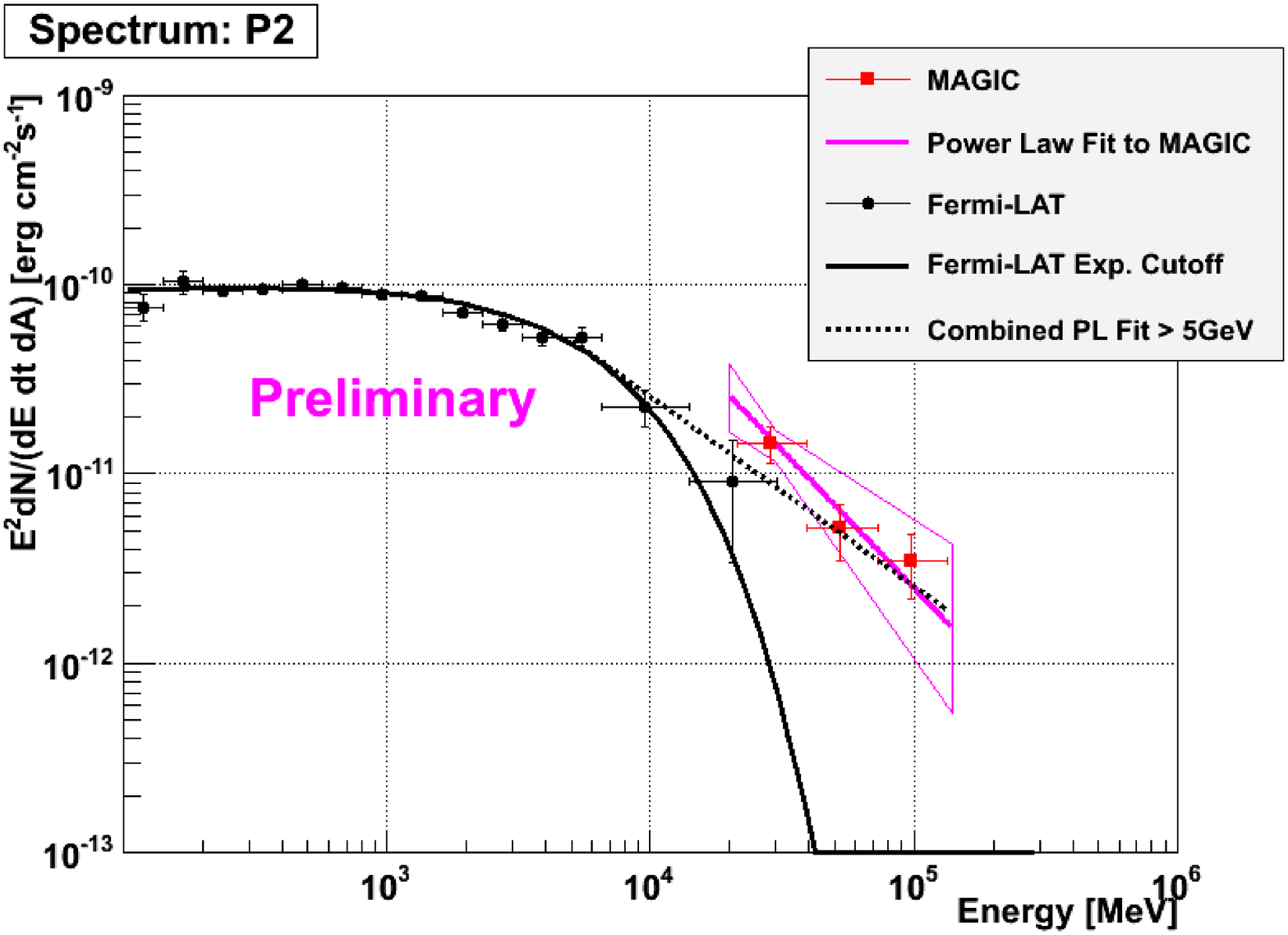}
\caption{The phase resolved energy spectra of the Crab pulsar measured
 by Fermi-LAT
and MAGIC with the Sum trigger (mono-mode). The energy spectra 
measured by Fermi-LAT is consistent with a power law with an exponential
cutoff, while the MAGIC measurements above 25 GeV deviate
from their extrapolation. The inconsistencies amount to 6.7$\sigma$, 3.0$\sigma$
and 5.8$\sigma$ level for P1~+~P2, P1 and P2, respectively.
} \label{FigSpectrum}
\end{figure*}

Based on these excess events, the phase resolved energy spectra of the Crab pulsar above 25 GeV were computed as shown in Fig.~\ref{FigSpectrum}. They can be well described by power laws and the best fit parameters are summarized in Table \ref{TabSpec}.
The energy spectrum measured by Fermi-LAT is also shown in the same figure.
For the Fermi-LAT points, 1 year of the Fermi-LAT data (from August 2008
to August 2009) were used. The best fit parameters obtained by the unbinned
likelihood analysis are summarized in Table \ref{TabSpec}.
The continuation from the Fermi-LAT measurements to the MAGIC measurements is rather smooth, 
while it is clear that the exponential cutoff spectrum determined by Fermi-LAT
is not valid in MAGIC energies. A detail statistical analysis showed 
the inconsistency amounts to 6.7$\sigma$, 3.0$\sigma$, and 5.8$\sigma$
for P1~+~P2, P1 and P2, respectively.
Further details of the results of the mono-mode observations 
are presented in \cite{TakaThesis}, \cite{MonoApJ} and \cite{MonoAstroph}.

\section{Stereo-mode observations}
As mentioned before, MAGIC observed the Crab pulsar with the stereo-mode since 2009.
Though the energy threshold of the stereo-mode (50 GeV) is higher than the mono-observation
with the Sum trigger (25 GeV), the VHE tail of the spectrum allows us to
detect the Crab pulsar above 50 GeV. The advantage of the stereo-mode observation
is higher background rejection power, better angular resolution and better energy resolution.
In the case of the stereo-mode observations, by using images from both telescopes, 
one can reconstruct the arrival direction of the gamma-rays better than in an ALPHA analysis.
The estimation of the shower maximum height is also more precise than 
in the mono-mode, which leads to the higher background rejection and the better energy resolution.

\begin{figure*}[h]
\centering
\includegraphics[width=2.in]{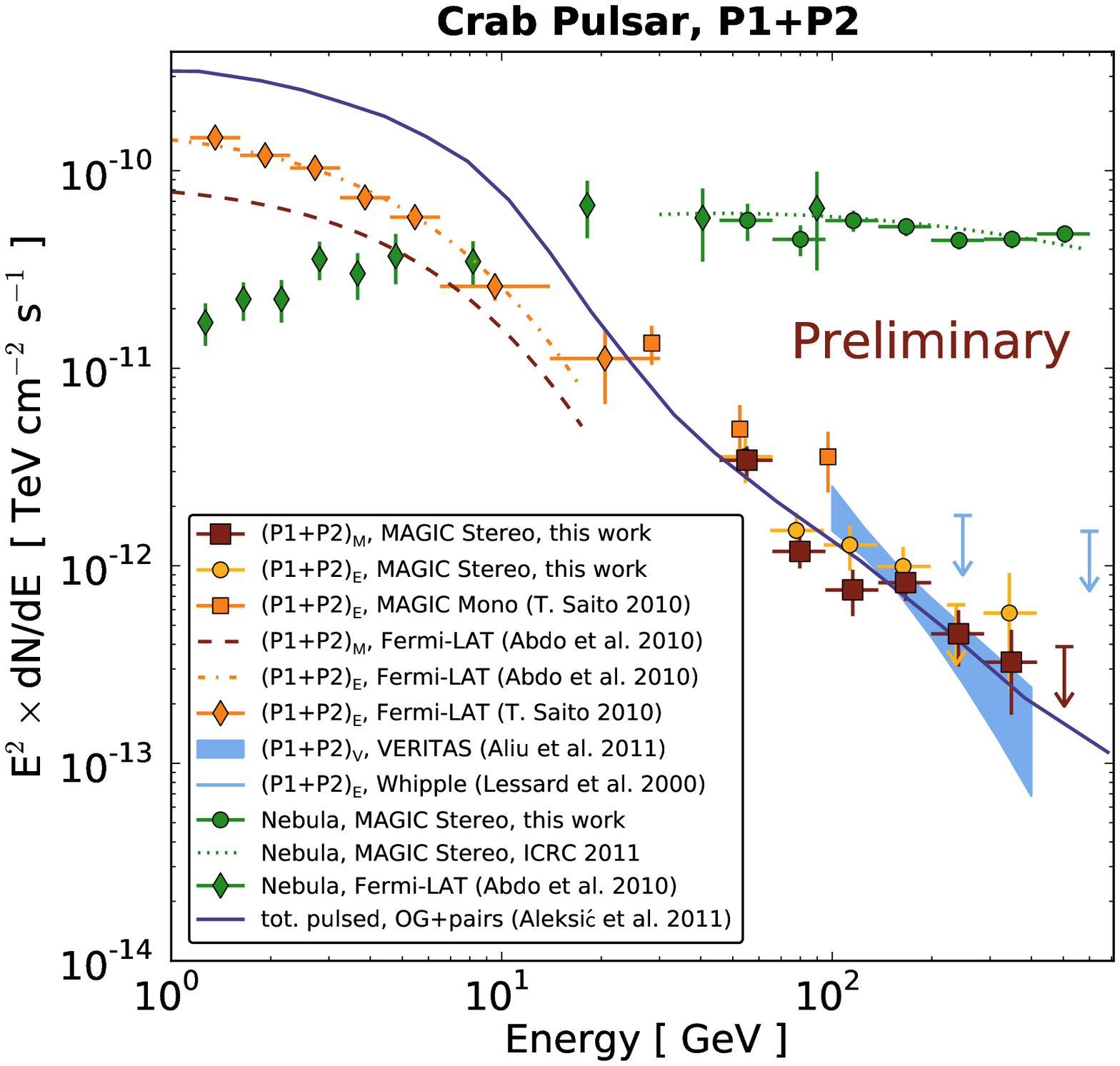}
\includegraphics[width=2.in]{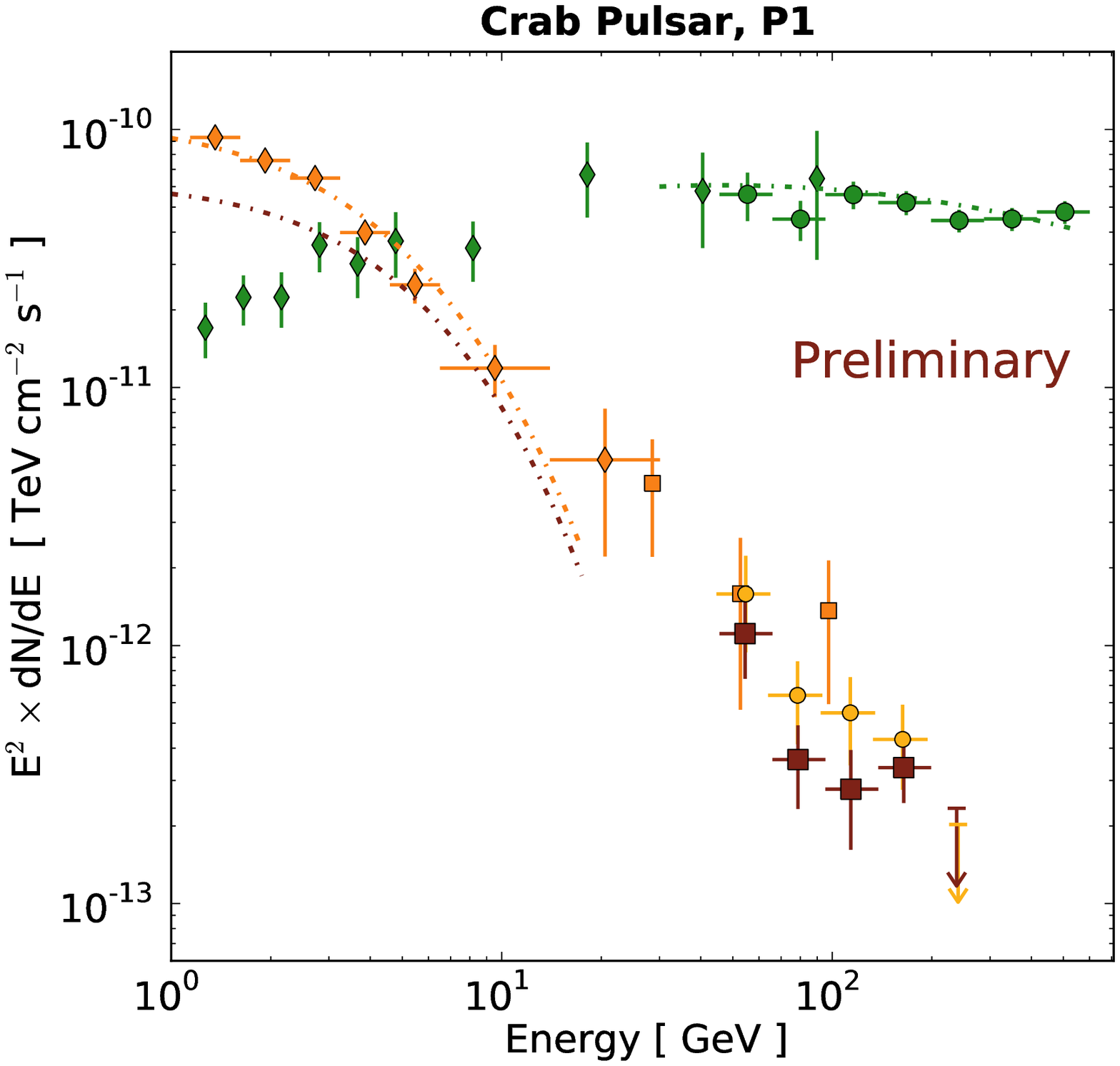}
\includegraphics[width=2.in]{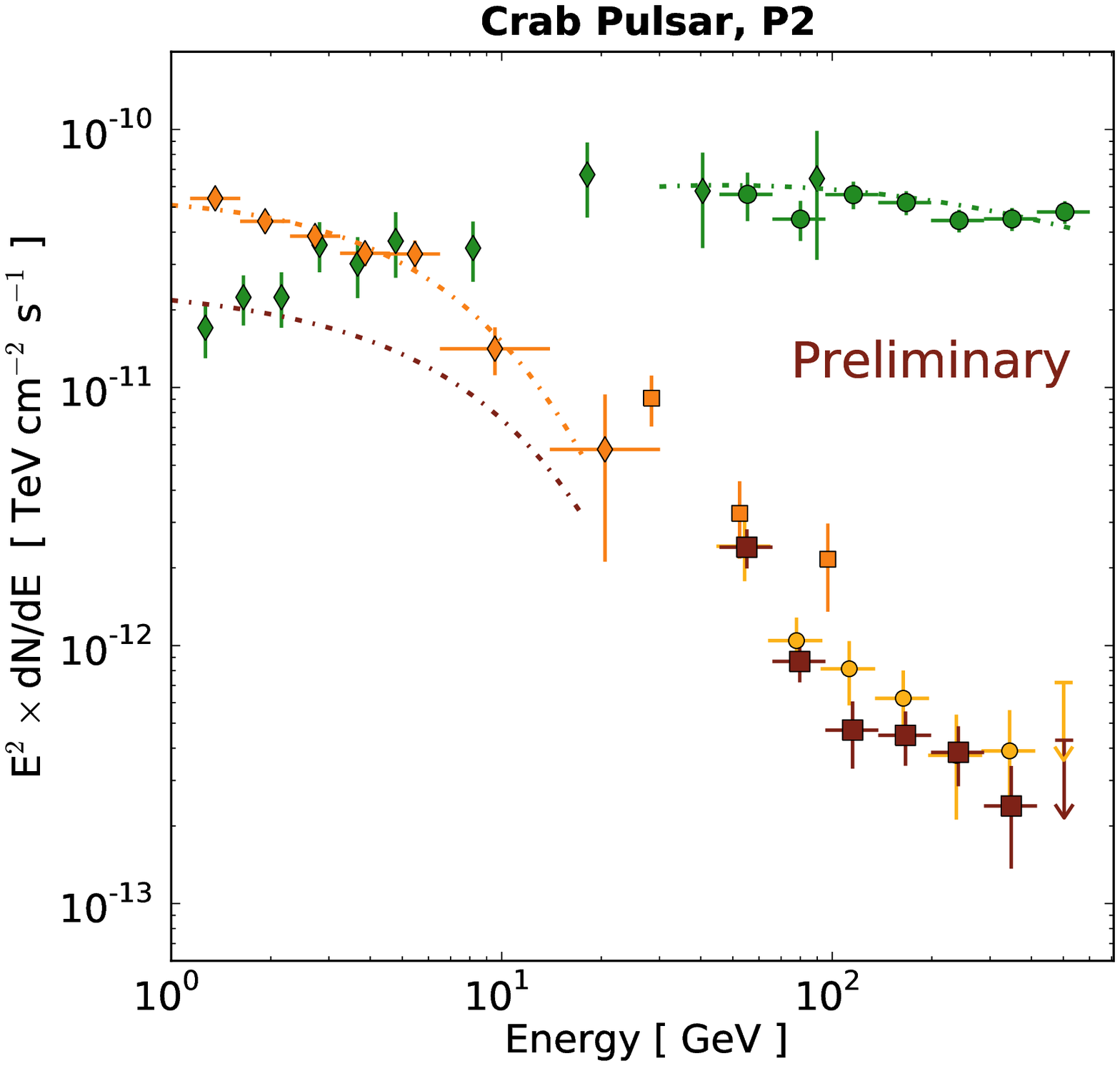}
\caption{The phase resolved energy spectrum of the Crab pulsar measured by Fermi-LAT, MAGIC-mono and MAGIC-stereo. After the break at around 5 GeV, 
the spectra follow power laws. (For the difference between the yellow points
and dark red points, see text.)
} \label{FigStereoSpectrum}
\end{figure*}

The lower panel of Fig. \ref{FigMonoLC} shows the light curve of the Crab pulsar obtained by 
stereo-mode observations. The total observation time used in this analysis is 73 hours.
Since the pulses are much narrower than the conventionally defined P1 and P2 phases,
the evaluation of the statistical significance of the excess was done by Z$^2_{10}$-test, H-test and $\chi^2$ test, which gave 8.6$\sigma$, 6.4$\sigma$ and 7.7$\sigma$, respectively.
By fitting two Gaussians to the two peaks, the peak positions are estimated to be
$0.005 \pm 0.003$ and $0.3996 \pm 0.0014$, while the corresponding FWHMs are $0.025 \pm 0.007$
and $0.026 \pm  0.004$. By defining the signal phases as $\pm 2 \sigma$ of the fitted Gaussian around the peaks,
the significance of the excess is 10.4$\sigma$, 5.5$\sigma$ and 9.9 $\sigma$ for P1~+~P2,
P1 and P2, respectively. The phase resolved energy spectra are also calculated and shown in Fig. \ref{FigStereoSpectrum}. The dark red squares denote
the spectra of the $\pm 2 \sigma$ phase interval described above,
while the yellow squares denote the ones with the conventional P1/P2 
definitions.
They smoothly connect with the mono-mode measurements and follow a power law. 
Further details of the results of the stereo-mode observations 
are presented in \cite{StereoAA}.

\section{Discussion and Conclusion}

The new measurements with MAGIC-mono and MAGIC-stereo
discovered that the energy spectrum of the Crab pulsar 
does not roll off as fast as the exponential cutoff but extends as a 
power law after the break at around 5 GeV. This suggests that the emission above 25 GeV is not dominated by
the curvature radiation as the standard OG and SG models predict.
Further theoretical studies and more observations are needed to understand 
the emission mechanism of the Crab pulsar. A theoretical interpretation
of the MAGIC measurements by K. Hirotani is presented in \cite{MonoApJ}, \cite{MonoAstroph} and \cite{StereoAA}.

\begin{table*}[h]
\centering
\begin{tabular}{|r|r|r|r|r|r|}
\hline
 & \multicolumn{3}{|c|}{Fermi-LAT} & \multicolumn{2}{|c|}{MAGIC-Mono}  \\
\hline
Phase & $f_1$ [10$^{-10}$cm$^{-2}$s$^{-1}$MeV$^{-1}$]& $\Gamma_1$ & $E_c$ & $f_{30  }$ 10$^{-9}$cm$^{-2}$s$^{-1}$TeV$^{-1}$] & $\Gamma_2$ \\
\hline
\hline
P1~+~P2 & $1.94 \pm 0.05$ & $1.98 \pm 0.02$ & $4.5 \pm 0.3$ & 14.9 $\pm$ 2.9 & 3.4 $\pm$ 0.5 \\
\hline	
P1      & $1.29 \pm 0.04$ & $1.99 \pm 0.02$ & $3.7 \pm 0.3$ &  4.5 $\pm$ 2.3 & 3.1 $\pm$ 1.0\\
 \hline
P2      & $0.67 \pm 0.02$ & $1.95 \pm 0.03$ & $5.9 \pm 0.7$ & $ 10.0 \pm 1.9$ & 3.4 $\pm$ 0.5 \\\hline
\end{tabular}
\centering
\footnotetext{In units of 10$^{-10}$cm$^{-2}$s$^{-1}$MeV$^{-1}$}
\footnotetext{In units of 10$^{-9}$cm$^{-2}$s$^{-1}$TeV$^{-1}$}
\caption{The best fit parameters of the spectra for different phase intervals.
The second to forth columns are obtained with Fermi-LAT data
assuming the spectral shape of $\frac{{\rm d}F(E)}{{\rm d} E} =  f_{1} (E/1{\rm ~[GeV]})^{-\Gamma_1}\exp(-E/E_c)$, while the fifth and sixth columns
are obtained with MAGIC-mono data
assuming the spectral shape of $\frac{{\rm d}F(E)}{{\rm d} E}= f_{30} (E/30{\rm ~[GeV]})^{-\Gamma_2} $
}
\label{TabSpec}
\end{table*}


\section*{Acknowledgments}

We would like to thank the Instituto db Astrof\'{\i}sica de
Canarias for the excellent working conditions at the
Observatorio del Roque de los Muchachos in La Palma.
The support of the German BMBF and MPG, the Italian INFN,
the Swiss National Fund SNF, and the Spanish MICINN is
gratefully acknowledged. This work was also supported by
the Marie Curie program, by the CPAN CSD2007-00042 and MultiDark
CSD2009-00064 projects of the Spanish Consolider-Ingenio 2010
programme, by grant DO02-353 of the Bulgarian NSF, by grant 127740 of
the Academy of Finland, by the YIP of the Helmholtz Gemeinschaft,
by the DFG Cluster of Excellence ``Origin and Structure of the
Universe'', by the DFG Collaborative Research Centers SFB823/C4 and SFB876/C3,
and by the Polish MNiSzW grant 745/N-HESS-MAGIC/2010/0.

\end{document}